\documentclass[aps,prl,twocolumn,amsmath,amssymb,showpacs,superscriptaddress,notitlepage]{revtex4-1}
\usepackage{graphicx}
\usepackage{subfigure}
\usepackage{dcolumn}
\usepackage{bm}
\usepackage{amsfonts}
\usepackage{mathrsfs}
\usepackage{amssymb}
\usepackage{amsmath}
\usepackage{color}
\usepackage{chngcntr}
\usepackage{stmaryrd}
\usepackage[colorlinks=true,breaklinks=true,linkcolor=blue,anchorcolor=red,citecolor=blue,urlcolor=blue]{hyperref}
\usepackage{booktabs}
\usepackage{multirow}
\usepackage{txfonts}
\usepackage[T1]{fontenc}

\def\bea{\begin{eqnarray}}
\def\eea{\end{eqnarray}}

\begin{document}

	\title{Unveiling non-Abelian statistics of vortex Majorana bound states in iron-based superconductors using fermionic modes}
	
	\author{Ming Gong}
\affiliation{International Center for Quantum Materials, School of Physics, Peking University, Beijing 100871, China}

	\author{Yijia Wu}
\affiliation{International Center for Quantum Materials, School of Physics, Peking University, Beijing 100871, China}

	\author{Hua Jiang}
\affiliation{School of Physical Science and Technology and Institute for Advanced Study, Soochow University, Suzhou 215006, China.}
	
\author{Jie Liu}
\affiliation{Department of Applied Physics, School of Science, Xian Jiaotong University, Xian 710049, China}

\author{X. C. Xie}
\email{xcxie@pku.edu.cn}
\affiliation{International Center for Quantum Materials, School of Physics, Peking University, Beijing 100871, China}
\affiliation{Beijing Academy of Quantum Information Sciences, Beijing 100193, China}
\affiliation{CAS Center for Excellence in Topological Quantum Computation,
	University of Chinese Academy of Sciences, Beijing 100190, China}

	\begin{abstract}
Motivated by the recent experiments that reported the discovery of vortex Majorana bound states (vMBSs) in iron-based superconductors, we establish a portable scheme to unveil the non-Abelian statistics of vMBSs using normal fermionic modes. The unique non-Abelian statistics of vMBSs is characterized by the charge flip signal of the fermions that can be easily read out through the charge sensing measurement. In particular, the charge flip signal will be significantly suppressed for strong hybridized vMBSs or trivial vortex modes, which efficiently identifies genuine vMBSs. To eliminate the error induced by the unnecessary dynamical evolution of the fermionic modes, we further propose a correction strategy by continually reversing the energy of the fermions, reminiscent of the quantum Zeno effect. Finally, we establish a feasible protocol to perform non-Abelian braiding operations on vMBSs.          
	\end{abstract}


	\maketitle

	\textcolor[rgb]{0.00,0.00,1.00}{\emph{Introduction}.}--Since the concept of quantum computation is proposed \cite{benioff_computer_1980,feynman_simulating_1982,feynman_quantum_1986}, decoherence, which stymies most of the realization approaches of quantum computers, becomes one of the thorniest challenges for this realm \cite{shor_scheme_1995,lidar_decoherence-free_1998}. By storing and operating quantum information non-locally, topological quantum computation (TQC) \cite{dennis_topological_2002,freedman_topological_2003,nayak_non-abelian_2008} evade this problem from the hardware level. Owning to the favored non-Abelian statistics, Majorana bound states are deemed as the most promising candidate for implementing TQC \cite{ivanov_non-abelian_2001,nayak_non-abelian_2008}. To date, a variety of schemes have been proposed to realize and manipulate such kind of quasi-particles in condensed matter systems, especially in topological superconductors (TSCs) \cite{fu_superconducting_2008,lutchyn_majorana_2010,cook_majorana_2011,hosur_majorana_2011,mi_proposal_2013,wang_chiral_2015,xu_topological_2016,pientka_topological_2017,liu_flux-induced_2019}. Among these, vortex Majorana bound states (vMBSs) \cite{hosur_majorana_2011,wang_topological_2015,xu_topological_2016,caroli_bound_1964,kong_half-integer_2019} are reported to be discovered recently in iron-based superconductors (FeSCs) such as FeTe$_{0.55}$Se$_{0.45}$ \cite{wang_topological_2015,zhang_observation_2018,wang_evidence_2018,chen_discrete_2018,chen_superconductivity_2018,chen_direct_2019,yang_vortex_2018,chen_observation_2020,zhang_observation_2021,liu_robust_2018,hanaguri_two_2018,hanaguri_quantum_2019,kong_half-integer_2019,kong_tunable_2020,zhu_nearly_2020,liu_new_2020}. These FeSCs integrate the advantages of high-$T_{c}$, topological band structure and self-proximity, making them highly promising in TQC \cite{shi_fete1xse_2017,hao_topological_2019,kong_emergent_2020}. 
	
    The first step toward the practical application of vMBSs in TQC is the demonstration of their non-Abelian statistics \cite{ivanov_non-abelian_2001}. Different from other proposals of realizing Majorana zero modes such as using semiconductor superconducting nanowires \cite{lutchyn_majorana_2010,lutchyn_majorana_2018}, vMBSs in FeSCs are tightly embedded into the Abrikosov lattice \cite{machida_zero-energy_2019,franz_quasiparticles_2000}, which complicates the fabrication of external structures and the implementation of non-Abelian braiding procedures. So far, experimental proposals for performing braiding operations on vMBSs mainly focus on moving the positions of vortices \cite{posske_vortex_2020,ma_braiding_2020,ma_braiding_2020-1}, which may be destructive to vMBSs and make the operation duration exceed the coherence time \cite{rainis_majorana_2012,budich_failure_2012}. Furthermore, these braiding schemes also make it difficult to reflect the non-Abelian statistics of vMBSs onto an experimental observable.     

\begin{figure}[t]
	\includegraphics[width=0.46\textwidth]{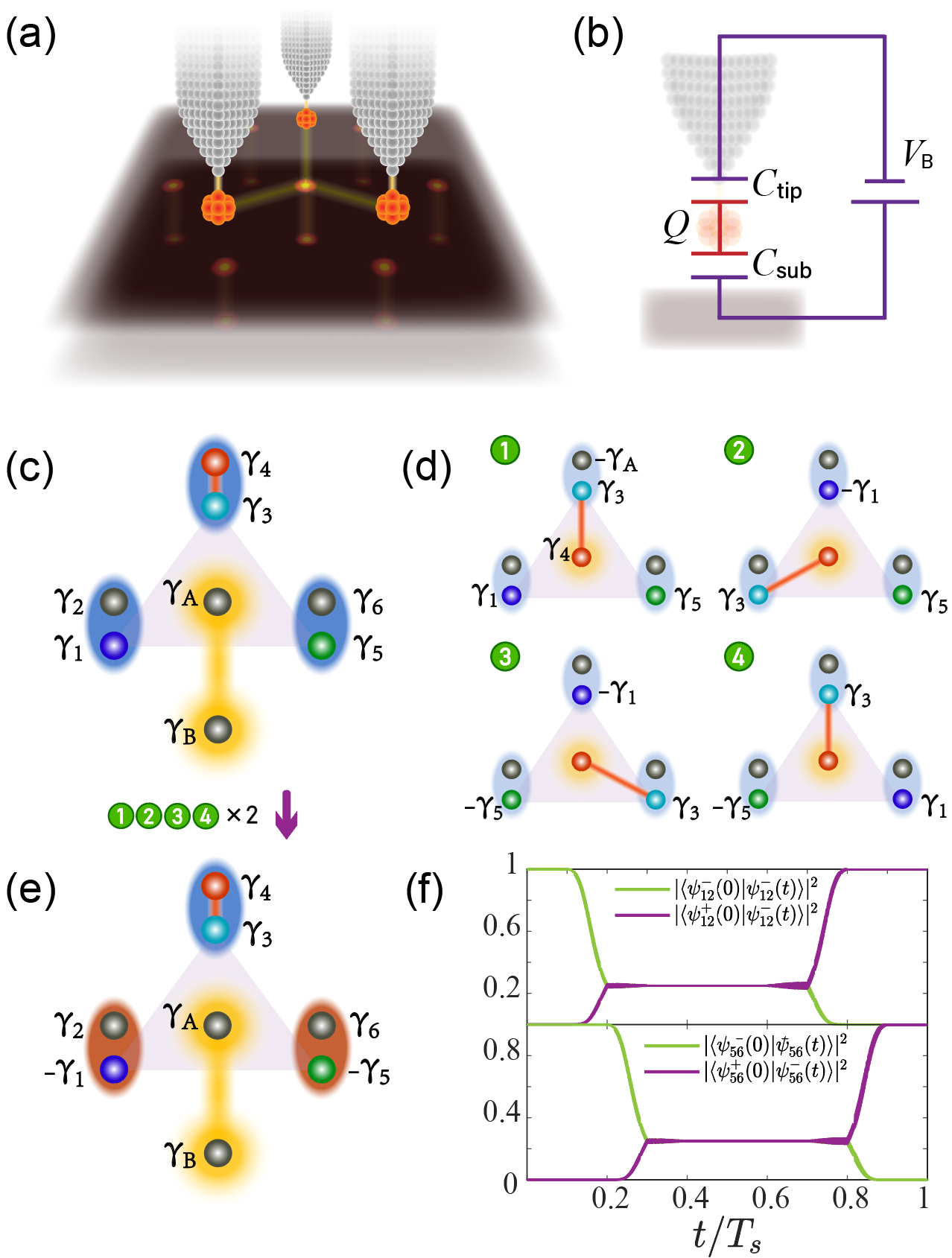}
	\caption{(a) Sketch of the fermionic Y-junction in FeSCs. (b) Illustration of the readout protocol for the CFS based on the charge sensing measurement.  (c) Minimal model for the fermionic Y-junction. $\gamma_{i}$ ($i=1,2,\dots,6$) represent the Majorana components of the fermions. $\gamma_{A}$ and $\gamma_{B}$ represent the vMBSs. (d) By alternately coupling the three fermionic modes to the vMBS $\gamma_{A}$, Majorana modes $\gamma_{1}$ and $\gamma_{5}$ undergo a non-Abelian braiding process. (e) Performing the braiding operation twice, the fermionic state $|\psi_{12}^{-}\rangle$ encoded by $\gamma_{1}$ and $\gamma_{2}$ filps to $|\psi_{12}^{+}\rangle$ and vice versa, same for $|\psi_{56}^{-}\rangle$. (f) Numerical results simulated in the TSC system. $T_{s}$ = 1${\rm \mu s}$ denotes the operation duration.} \label{fig1}
\end{figure}

	In this Letter, we establish a portable scheme to unveil the non-Abelian statistics of vMBSs in FeSCs using normal fermionic modes (see Fig.~\ref{fig1}). 
    By alternately coupling fermionic modes to the vMBS, Majorana components of the fermions undergo a non-Abelian braiding process, resulting in the charge flip signal (CFS) of the fermions and greatly simplifies the readout protocol through charge sensing measurements \cite{munk_parity--charge_2020,szechenyi_parity--charge_2020}.  Moreover, the CFS is significantly suppressed when the vMBSs are strongly hybridized or the vortex bound state is fermionic \cite{prada_andreev_2020}. 
	For this reason, it provides a feasible method to distinguish vMBSs from trivial Andreev bound states. Experimentally, our proposal can be conveniently realized in FeSCs with the help of AFM/STM tips \cite{krieg_atomic_2019,yin_probing_2021,ma_atomic_2020}.  
	To improve the quality of the CFS, error induced by unnecessary dynamical evolution of the fermionic modes should be eliminated. 
	We propose a method to correct such a dynamical error by frequently reversing the energy of the fermions. Such an operation freezes the dynamical evolution of low-energy modes and can be understood as a Majorana version of quantum Zeno effect \cite{misra_zenos_1977}. 
	In experiments, the above reversing process can be achieved through spin-echo-like techniques \cite{jones_geometric_2000}. 
	Finally, using a single fermionic mode, we propose a portable protocol to perform the braiding operations over vMBSs. The braiding completeness is closely related to the geometric phase of Majorana modes accumulated during the braiding process \cite{liu_minimal_2021}. 
	Our proposals shed light on scalable TQC in FeSCs.     

	\textcolor[rgb]{0.00,0.00,1.00}{\emph{Basic setup: the fermionic Y-junction.}}--We establish a fermionic Y-junction setup, which consists of three fermionic modes ($\psi_{12}$, $\psi_{34}$, and $\psi_{56}$) and a pair of vMBSs ($\gamma_{A}$ and $\gamma_{B}$). Here, $\psi_{ij}$ ($i,j=1,2,\dots,6$) denotes the annihilation operator of the fermionic mode with Majorana components $\gamma_{i}$ and $\gamma_{j}$ [i.e. $\psi_{ij}=(\gamma_i+i\gamma_j)/2$ and $\psi_{ij}^{\dagger}=(\gamma_i-i\gamma_j)/2$]. As sketched in Fig.~\ref{fig1} (c), the minimal model Hamiltonian is
	\begin{eqnarray}
		\nonumber
	H_{Y}&&=iE_{d,12}\gamma_{1}\gamma_{2}/2+iE_{d,34}\gamma_{3}\gamma_{4}/2+iE_{d,56}\gamma_{5}\gamma_{6}/2\\
	&&+it_{A,1}\gamma_{A}\gamma_{1}/2+it_{A,3}\gamma_{A}\gamma_{3}/2+it_{A,5}\gamma_{A}\gamma_{5}/2,
	\end{eqnarray}
	 where $E_{d,ij}$ denotes the energy of $\psi_{ij}$, while $t_{A,i}$ is the coupling strength between the vMBS $\gamma_{A}$ and the Majorana mode $\gamma_{i}$ inside the fermion. Similar to a traditional Y-junction \cite{karzig_universal_2016}, by alternately turning on and off $t_{A,i}$, Majorana components of the fermions are transmitted spatially. Detailed operation procedure is as follows. Firstly, parameters in $H_{Y}$ is initialized as $E_{d,34}=E_{0}$ with $E_{d,12}=E_{d,56}=t_{A,i}=0$. Under this condition, $\gamma_{3}$ and $\gamma_{4}$ are in a strong-coupled status, with all the other Majorana modes frozen at zero energy. 
	 In step 1, we gradually turn off $E_{d,34}$ and turn on $t_{A,3}$ from 0 to $t_{c}$. By doing so, $\gamma_{4}$ ($\gamma_{A}$) is transmitted to the original position of $\gamma_{A}$ ($\gamma_{4}$) with $\gamma_{A}$ picking up a minus sign due to the non-Abelian statistics.  Similar to step 1, the next three steps and the resulting configurations of Majorana modes are illustrated in Fig.~\ref{fig1} (d). In the final step, we turn off $t_{A,3}$ and turn on $E_{d,34}$ so that $H_{Y}$ comes back to its initial form, preparing for the next cycle of operation. After performing the above steps, $\gamma_{1}$ and $\gamma_{5}$ undergo a non-Abelian braiding process with all the other Majorana modes go back to their initial positions. Performing twice, both $\gamma_{1}$ and $\gamma_{5}$ pick up minus signs as shown in Fig.~\ref{fig1} (e). From the viewpoint of the fermionic modes, $\psi_{12}=(\gamma_1+i\gamma_2)/2$ flips to $-\psi_{12}^{\dagger}=(-\gamma_1+i\gamma_2)/2$ with the corresponding state $|\psi_{12}^{+}\rangle$ flipping to $|\psi_{12}^{-}\rangle$ and vice versa, resulting in the CFS ($|\psi_{ij}^{\pm}\rangle$ represents the state encoded by $\gamma_{i}$ and $\gamma_{j}$, where the superscripts $+$ and $-$ denote the occupation and unoccupation state of the fermionic mode). Same results also apply to $|\psi_{56}^{\pm}\rangle$. 
	 
	 We numerically simulate the above process in a two-dimensional TSC system that mimic the surface TSC emerged in FeSCs \cite{pathak_majorana_2021-1}. The lattice Hamiltonian is
	 	\begin{eqnarray}
	 	\scriptsize
	 	\nonumber
	 	H_{{\rm TSC}}=&&\sum_{\mathbf{i}}\Big[\frac{i \hbar v_{F}}{2 a}(c_{\mathbf{i}}^{\dagger} \sigma_{y} c_{\mathbf{i}+\delta \hat{\mathbf{x}}}-c_{\mathbf{i}}^{\dagger} \sigma_{x} c_{\mathbf{i}+\delta \hat{\mathbf{y}}})-\frac{\mu}{2}c_{\mathbf{i}}^{\dagger}\sigma_{0} c_{\mathbf{i}}\nonumber\\
	 	&&-\frac{W}{2a}(c_{\mathbf{i}}^{\dagger}\sigma_{z} c_{\mathbf{i}+\delta\hat{\mathbf{x}}}+c_{\mathbf{i}}^{\dagger} \sigma_{z}c_{\mathbf{i}+\delta \hat{\mathbf{y}}})+\frac{W}{a}c_{\mathbf{i}}^{\dagger}\sigma_{z} c_{\mathbf{i}}\nonumber\\
	 	&&+\Delta(\mathbf{i})c_{\mathbf{i},\uparrow}^{\dagger}c_{\mathbf{i},\downarrow}^{\dagger}\Big]+\mathrm{H.c.}
	 	\label{eq:1},
	 	\end{eqnarray}
	 where the first term represents the topological surface states with $v_{F}$ the fermi velocity and $a$ the lattice constant. The fermion doubling problem is eliminated by adding a Wilson mass term with strength $W$  \cite{marchand_lattice_2012,gong_transport_2020,pathak_majorana_2021-1}. $\mu$ denotes the chemical potential. TSC emerges by adding s-wave pairing terms with pairing potential $\Delta$  \cite{fu_superconducting_2008}. Vortices are introduced through $\Delta(\mathbf{i})=\Delta {\rm tanh}\frac{|\mathbf{i}-\mathbf{j}|}{\xi}e^{i\theta(\mathbf{i}-\mathbf{j})}$ where $\mathbf{j}$ denotes the location of the vortex core, $\xi$ is the coherence length and $\theta(\mathbf{i}-\mathbf{j})$ is the superconducting phase. Here, we consider a vortex and an anti-vortex that support vMBSs $\gamma_{A}$ and $\gamma_{B}$ for simplicity. The fermionic modes are spin-polarized \cite{hu_theory_2016} and couple to vMBS $\gamma_{A}$ with coupling strength $t_{A,i}$ ($i=1,3,5$). During the operation, $t_{A,i}$'s are alternately turned on and off ranging from 0 to $t_{c}$. The  simulation parameters are taken as $a=1$, $\hbar v_{F}=1$, $W=1$, $\mu=0$, $\Delta=1.5$, $\xi=2$, $E_{0}=0.3$, and $t_{c}=0.1$. The total operation duration $T_{s}=1.53\times10^{6}$, corresponding to 1 ${\rm \mu s}$ (with the energy unit meV) in SI units. Specially, the adiabatic condition $\hbar/T_{s}\ll E_{c}$ should be satisfied in experiments ($E_{c}$ denotes the lowest excitation energy above the vMBSs that could be the energy of the superconducting gap or the lowest sub-gap CdGM state). For FeSCs such as FeTe$_{0.55}$Se$_{0.45}$, LiFeAs, and CaKFe$_{4}$As$_{4}$, the observed energy scale of $E_{c}$ are of the order meV \cite{zhang_observation_2018,kong_half-integer_2019,kong_tunable_2020,liu_new_2020}, hence $T_{s}\gg \hbar/(1{\rm meV})=6.56\times 10^{-7}{\rm \mu s}$ is easily satisfied for microsecond scale operations. Fermionic states are initialized to $|\psi^{-}_{12}\rangle$ and $|\psi^{-}_{56}\rangle$. $|\psi^{-}_{ij}(t)\rangle=U(t)|\psi^{-}_{ij}(0)\rangle$ where $U(t)=\hat T{\rm exp}[-i\int_{0}^{t}d\tau H(\tau)]$ is the time evolution operator ($\hat T$ is the time-ordering operator)  \cite{amorim_majorana_2015,wu_non-abelian_2020}. The resulting transition probabilities $|\langle \psi_{12}^{+} (0)|\psi_{12}^{-}(t)\rangle|^{2}$ and $|\langle \psi_{56}^{+} (0)|\psi_{56}^{-}(t)\rangle|^{2}$ can serve as the CFSs, which manifest the flip of the fermion state from the unoccupied $|\psi^{-}_{12}\rangle$ ($|\psi^{-}_{56}\rangle$) to occupied $|\psi^{+}_{12}\rangle$ ($|\psi^{+}_{56}\rangle$), as demonstrated in Fig.~\ref{fig1} (f). 
	 
	 FeSCs is an ideal platform to realize the fermionic Y-junction. The fermionic mode could be achieved through quantum-dots, molecular clusters or other confined nano-structures  \cite{beenakker_theory_1991,park_coulomb_2002}. By attaching these structures on top of AFM/STM tips  \cite{krieg_atomic_2019,yin_probing_2021} and driving them to approach the vortex core in turn [Fig.~\ref{fig1} (a)], the coupling parameters $t_{A,i}$ ($i=1,3,5$) change alternately. Consequently, Majorana components of fermion states undergo non-Abelian braiding process, leading to CFS in these nano-structures which can be detected through charge sensing measurements \cite{johnson_coulomb-modified_2004,berkovits_theory_2005,miyahara_quantum_2017,leoni_controlling_2011} [see Fig.~\ref{fig1} (b)].               	
	\begin{figure}[t]
	\includegraphics[width=0.48\textwidth]{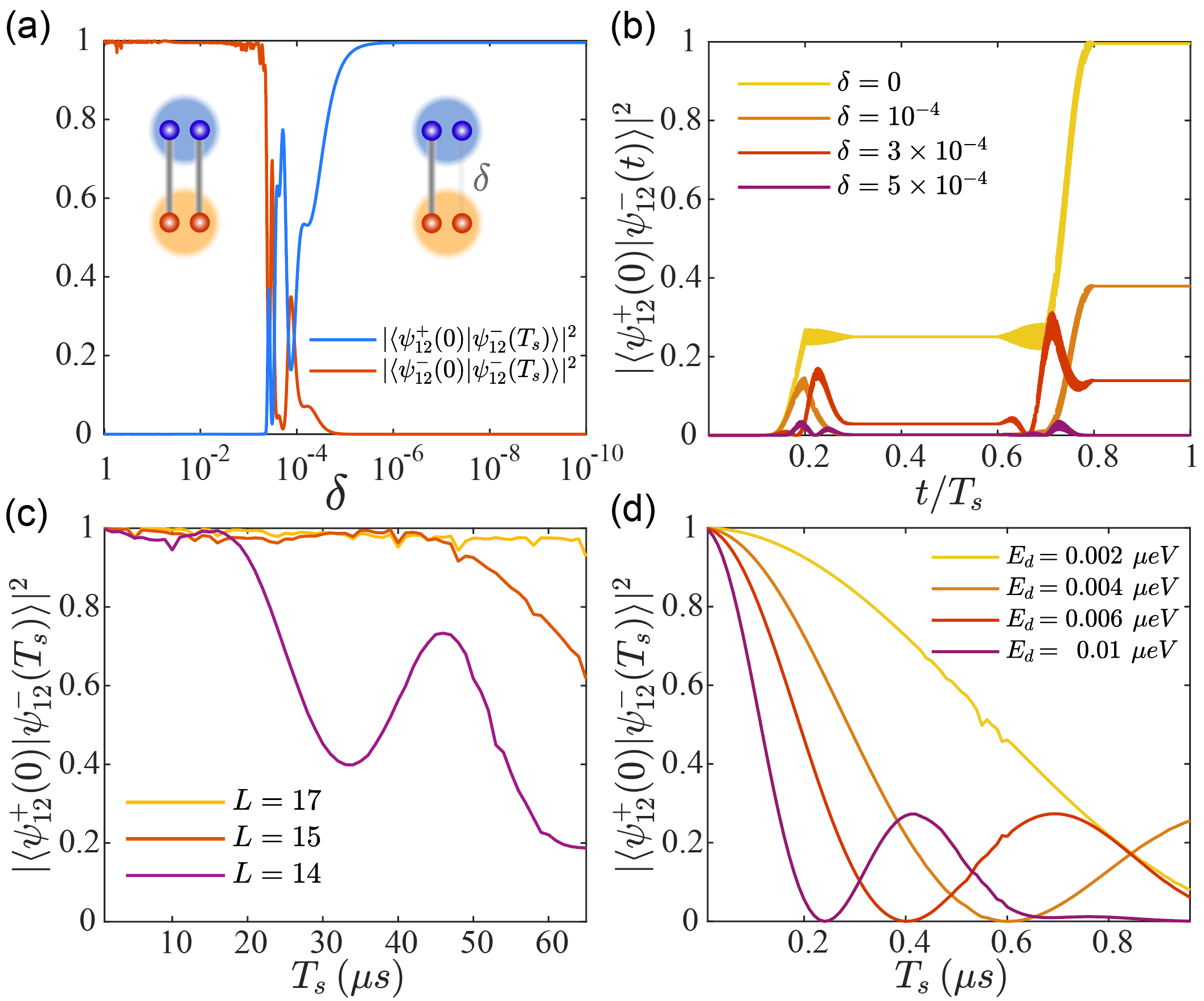}
	\caption{(a) Numerical result of the CFS for $H_{Y,\delta}$ versus the controlling parameter $\delta$. $T_{s}$=1${\rm \mu s}$, same for (b). (b) Flip probability from unoccupied state $|\psi_{12}^{-}\rangle$ to the occupied state $|\psi_{12}^{+}\rangle$ during the operation under different $\delta$. (c) CFS versus the operation duration $T_{s}$ with different distance $L$ between vortices. (d) CFS versus $T_{s}$ with different $E_{d}$ for the fermionic modes.}  
	\label{fig2}
    \end{figure}
	
	\textcolor[rgb]{0.00,0.00,1.00}{\emph{Identifying vMBSs using the fermionic Y-junction.}}--
    In sharp contrast to the traditional Y-junction that contains only Majorana modes, the CFS of our fermionic Y-junction is highly dependent on the Majorana nature of the vortex mode. The CFS will be destructed by replacing the vMBS into a fermionic mode. 
	To demonstrate such a consequence, we replace $H_{Y}$ to $H_{Y,\delta}=H_{Y}+h_{Y,\delta}$ with $
    h_{Y,\delta}=i\delta t_{A,1}\gamma_{B}\gamma_{2}/2+i\delta t_{A,3}\gamma_{B}\gamma_{4}/2+i\delta t_{A,5}\gamma_{B}\gamma_{6}/2$, where $\delta$ is a controlling parameter varying from 0 to 1. $\delta=0$ corresponds to the case where the vMBS only couples to half of a fermionic mode \cite{flensberg_non-abelian_2011}, and we call it the Majorana-type coupling. The case $\delta=1$ represents a fermion-type coupling between the fermionic mode $\psi_{AB}$ (encoded by $\gamma_{A}$ and $\gamma_{B}$) and $\psi_{ij}$.
    As demonstrated in Fig.~\ref{fig2} (a) and (b), the CFS is significantly suppressed by increasing $\delta$ from 0 to 1, which corresponds to a transition from the Majorana-type coupling to the fermion-type coupling \cite{prada_andreev_2020}.  In FeSCs, the vortex mode may be a normal Andreev bound state which is a fermionic excitation  \cite{caroli_bound_1964,yin_observation_2015,jiang_quantum_2019,chen_non-abelian_2020} that results in the absence of the CFS. For this reason, our fermionic Y-junction can serve as a detector to distinguish vMBSs from other trivial states. Moreover, the CFS can also be suppressed when the hybridization between vMBSs becomes stronger  \cite{cheng_tunneling_2010,cheng_splitting_2009,chiu_scalable_2020}. When the vortices get closer [Fig.~\ref{fig2} (c)], the CFS oscillates to zero as $T_{s}$ increases. For hybridized vMBSs, the fermion modes in the junction not only couple to the nearest vMBS $\gamma_{A}$, but also partially couple to $\gamma_{B}$ in the distance, resulting in a non-zero $\delta$ in $H_{Y,\delta}$ and destructs the CFS. Therefore, our fermionic Y-junction can also help to pick out genuine vMBSs for TQC in FeSCs.

\begin{figure}[t]
	\includegraphics[width=0.48\textwidth]{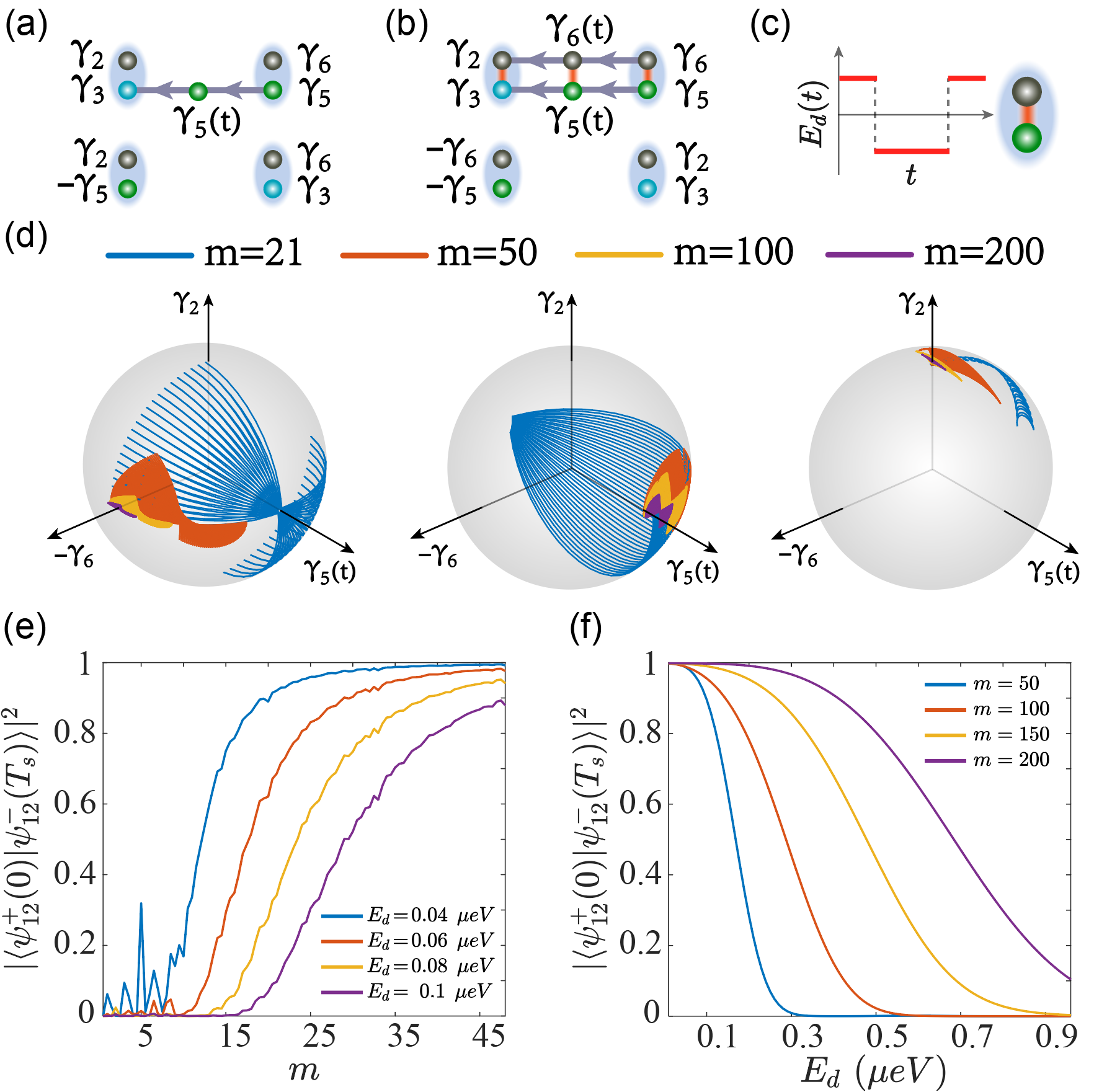}
	\caption{Sketch of step 3 in Fig.~\ref{fig1} without (a) and with (b) dynamical error. (c) The error-correcting strategy by continually reversing $E_{d}$. (d) Illustration of the Majorana version of quantum Zeno effect. The trajectories of $\gamma_{6}$, $\gamma_{2}$, and $\gamma_{5}(t)$ under dynamical evolution sweep across the Bloch sphere when the reversing frequency is small ($m$=21) but are frozen at the original positions as it increases ($m$=200).  (e) [(f)] Numerical results of CFS as a function of $m$ ($E_{d}$)  under different $E_{d}$ ($m$) simulated in the TSC system. The time duration $T_{s}$=1${\rm \mu s}$.} \label{fig3}
\end{figure}
	\textcolor[rgb]{0.00,0.00,1.00}{\emph{Correction of the dynamical error.}}--
     So far, we have assumed that the energy of the fermionic modes are fixed at 0 during the operation. However, in real nanostructures, the deviation of the on-site energy of fermionic states is inevitable, which brings error by enabling the dynamical evolution of low energy states that suppress the non-Abelian braiding process of Majorana modes ($\gamma_{1}$ and $\gamma_{5}$). We numerically demonstrate such a dynamical error in the TSC system. Here, we set $E_{d,12}=E_{d,56}=\tilde{E}_{d,34}=E_{d}$ \footnote{Here, we take $E_{d,12}=E_{d,56}=\tilde{E}_{d,34}=E_{d}$ for two reasons: \romannumeral1) The relative signs for $E_{d,12}$, $E_{d,56}$, and $\tilde{E}_{d,34}$ are not important under the particle-hole transformations. \romannumeral2) The dynamical error is dominated by the evolution of the fermionic mode with the largest energy $E_{d,{\rm max}}={\rm max}\{E_{d,12},E_{d,56},\tilde{E}_{d,34}\}$. We investigate the worst case with $E_{d,12}=E_{d,56}=\tilde{E}_{d,34}=E_{d,{\rm max}}$ for the simplicity in analysis.} where $\tilde{E}_{d,34}$ is the minimal value of $E_{d,34}$. As shown in Fig.~\ref{fig2} (d), the increasing of $E_{d}$ makes the CFS drops dramatically and narrows $T_{s}$ into a very small scale (typically 0.1 ${\rm \mu s}$). 
     
     The dynamical error originates from the evolution of low-energy states. We take one typical step [step 3 in Fig.~\ref{fig1} (d)] as an example to illustrate its mechanism. As compared in Fig.~\ref{fig3} (a) and (b), dynamical effect becomes significant when $E_{d}$ deviates from 0, resulting in an additional exchange process between $\gamma_{2}$ and $\gamma_{6}$, thus causes error to the CFS. We propose a error-correcting strategy by continually reversing $E_{d}$ as $E_{d}(t)=E_{d}{\rm sign} \left[ {\rm cos}\frac{2\pi m t}{T} \right]$ ($m$ determines the reversing frequency) [Fig.~\ref{fig3} (c)]. As $m$ increases the non-Abelian braiding process as well as the CFS will be recovered. We use the Hamiltonian $
	H_{\rm eff}=\frac{iE_{d}(t)}{2}(\gamma_{3}\gamma_{2}+\gamma_{5}\gamma_{6})+\frac{it_{c}}{2}\gamma_{4}[\gamma_{3}{\rm cos}\theta(t)+\gamma_{5}{\rm sin}\theta(t)]$
	to model such a process. Here, $\theta(t)=\frac{\pi t}{2T}$ controls the relative coupling strength, and $T$ is the time duration. 
	Under the transformations:  $\gamma_{2}(t)=\gamma_{2}{\rm cos}\theta(t)+\gamma_{6}{\rm sin}\theta(t)$, $
	\gamma_{3}(t)=\gamma_{3}{\rm cos}\theta(t)+\gamma_{5}{\rm sin}\theta(t)$, $
	\gamma_{5}(t)=\gamma_{5}{\rm cos}\theta(t)-\gamma_{3}{\rm sin}\theta(t)$, and $
	\gamma_{6}(t)=\gamma_{6}{\rm cos}\theta(t)-\gamma_{2}{\rm sin}\theta(t)$, $
	H_{\rm eff}$ can be rewritten as
	\begin{eqnarray} H_{\rm eff}=\frac{iE_{d}(t)}{2}\gamma_{5}(t)\gamma_{6}(t)+i\gamma_{3}(t)[\frac{E_{d}(t)}{2}\gamma_{2}(t)-\frac{t_{c}}{2}\gamma_{4}].
	\label{Heff}
	\end{eqnarray} 
	As $t_{c}/E_{d}(t)\rightarrow \infty$, the last term in Eq.~(\ref{Heff}) that describe the high energy $\gamma_{3}(t)$ and $\gamma_{4}$ play little role to generate low-energy dynamics. Therefore, dynamical error becomes dominant only in the subspace spanned by $\gamma_{2}$, $\gamma_{6}$, and $\gamma_{5}(t)$.
	Adopting the relation $e^{\alpha \gamma_{1}\gamma_{2}}\gamma_{1}e^{-\alpha \gamma_{1}\gamma_{2}}={\rm cos}2\alpha \gamma_{1}-{\rm sin}2\alpha \gamma_{2}$, the time evolution operator 	$U(T)=\hat{T}\int_{0}^{T}e^{ \frac{E_{d}(\tau)}{2}\gamma_{5}(\tau)\gamma_{6}(\tau)d\tau}$ can be approximated by successive rotations in the Euclidian space [$\hat x$(-$\gamma_{6}$), $\hat y$($\gamma_{5}(t)$), $\hat z$($\gamma_{2}$)] as ${\rm R}_{\hat n_{2m}}(\frac{-E_{d}T}{2m}){\rm R}_{\hat n_{2m-1}}(\frac{E_{d}T}{2m}) \dots {\rm R}_{\hat n_{2}}(\frac{-E_{d}T}{2m}){\rm R}_{\hat n_{1}}(\frac{E_{d}T}{2m})$ where ${\rm R}_{\hat n}(\phi)$ denotes the rotation around the axis $\hat n$ by $\phi$ and $\hat n_{k}={\rm cos}\frac{\pi k}{4m}\hat z -{\rm sin}\frac{\pi k}{4m}\hat x$ . The dynamical evolution of Majorana modes is gradually frozen by increasing the reversing frequency. For example, comparing $m$=21 and $m$=200 in Fig.~\ref{fig3} (d), $\gamma_{2}$, $\gamma_{6}$, and $\gamma_{5}(t)$ are pinned to their original positions for the larger $m$, thus successfully correct the dynamical error. Furthermore, as shown in Fig.~\ref{fig3} (e) and (f), the CFS (simulated in TSC system) is gradually recovered as $m$ increases. Interestingly, the physics behind our correction strategy is consistent with the well-known quantum Zeno effect that illustrates the stabilization of quantum states under frequent measurements or disturbance  \cite{misra_zenos_1977,syassen_strong_2008,berry_transitionless_2009,hu_majorana_2015}. Here, the quantum Zeno effect freezes both the dynamical and geometric evolutions of Majorana modes $\gamma_{2}$ and $\gamma_{6}$ as well as protects the adiabatic non-Abelian process between $\gamma_{3}$ and $\gamma_{5}$. In experiments, frequently reversing of the on-site energy of the fermionic mode can be realized through spin-echo-like schemes \cite{jones_geometric_2000}.

	\begin{figure}[t]
	\includegraphics[width=0.48\textwidth]{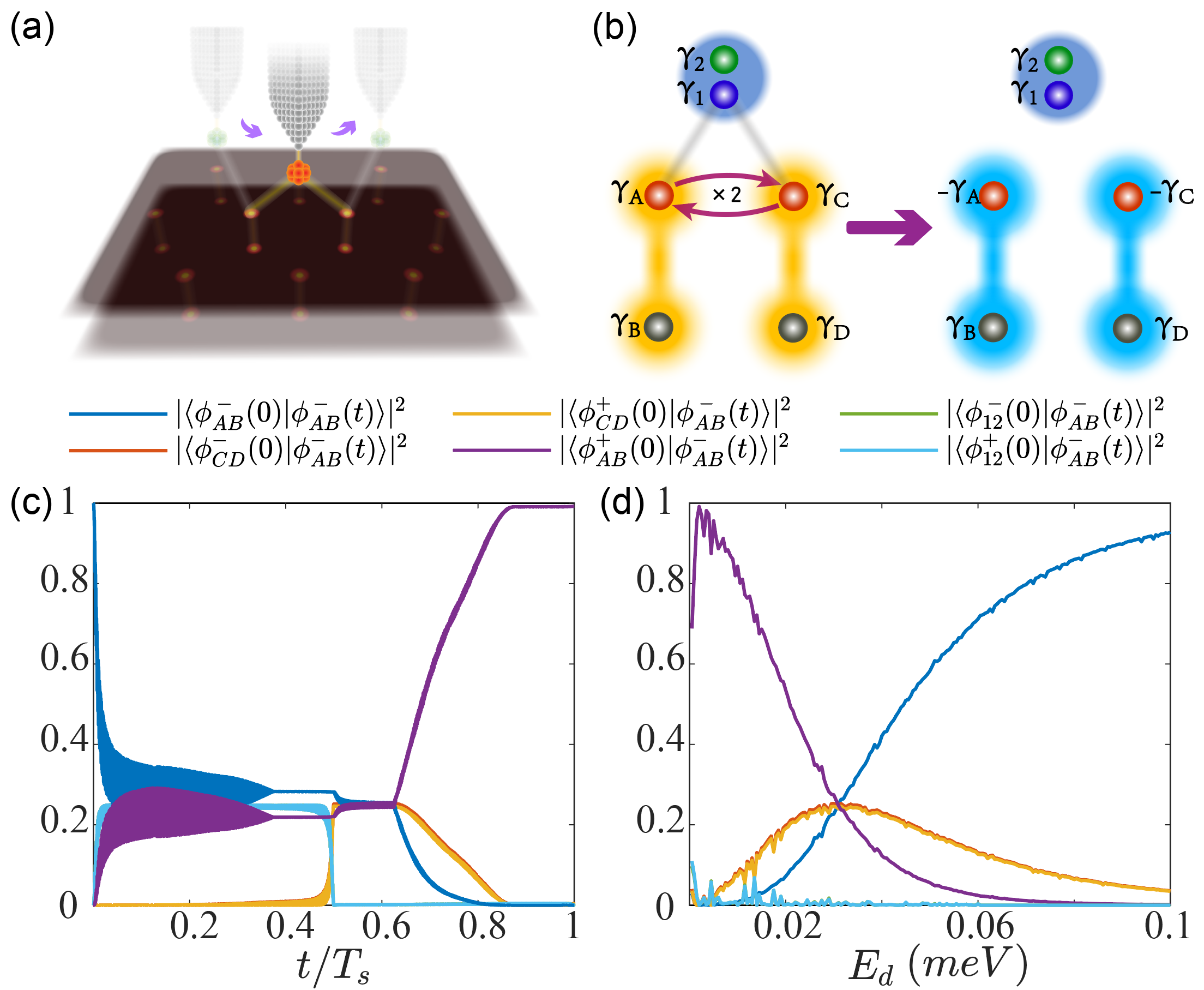}
	\caption{(a) Sketch of the non-Abelian braiding protocol for vMBSs using a single fermionic mode in FeSCs. (b) vMBSs $\gamma_{1}$ and $\gamma_{3}$ undergo non-Abelian braiding process, resulting in the flip of the qubits $|\phi_{12}^{\pm}\rangle$ and $|\phi_{34}^{\pm}\rangle$. (c) Transition probabilities during the braiding operation with $E_{d}=0.002$ and $t_{c}=0.05$. (d) Braiding completeness as a function of $E_{d}$ with $t_{c}=0.05$. The braiding duration $T_{s}$=$65.36$ ${\rm \mu s}$.} \label{fig4}
\end{figure}	
	
	\textcolor[rgb]{0.00,0.00,1.00}{\emph{Non-Abelian braiding protocol for vMBSs using a single fermionic mode.}}-- Based on the vMBSs that are identified by our fermionic Y-junction as weak-hybridized and possessing excellent non-Abelian statistical properties, we further propose a portable and scalable protocol to perform non-Abelian braiding operations over these vMBSs in FeSCs. The braiding operation is implemented by coupling a single fermionic mode to a pair of vMBSs alternately [Fig.~\ref{fig4} (a)]. Such a braiding protocol greatly simplifies the experimental setup and minimizes the damages to the vMBSs during the braiding process.
	
	As shown in Fig.~\ref{fig4} (b), the fermionic state $|\phi_{AB}^{-}\rangle$ ($|\phi_{CD}^{-}\rangle$) encoded by vMBSs $\gamma_{A}$ and $\gamma_{B}$ ($\gamma_{C}$ and $\gamma_{D}$) flips to $|\phi_{AB}^{+}\rangle$ ($|\phi_{CD}^{+}\rangle$), and vice versa . The transition probability $|\langle \phi_{AB}^{+}(0)|\phi_{AB}^{-}(T_{s})\rangle|^{2}$ (or $|\langle \phi_{CD}^{+}(0)|\phi_{CD}^{-}(T_{s})\rangle|^{2}$) signals the braiding completeness for our protocol. This quantity is directly related to the solid angle $\Omega_{c}={\rm arccos}(E_{d}/\sqrt{E_{d}^{2}+t_{c}^{2}})$ (which is also the geometric phase of $\gamma_{A}$ and $\gamma_{C}$ accumulated during the operation) enclosed by the trajectory of $\gamma_{2}$ in the space ($\gamma_{2}$, $\gamma_{A}$, $\gamma_{C}$) through the relation $|\langle \phi_{CD}^{+}(0)|\phi_{CD}^{-}(T_{s})\rangle|=\frac{1-{\rm cos}(2\Omega_{c})}{2}=\frac{t_{c}^{2}}{E_{d}^{2}+t_{c}^{2}}$ \cite{liu_minimal_2021}. 	
	Numerical simulations of our braiding protocol with two pairs of vortices (see details in the Supplementary material) demonstrates that the vMBSs qubit successfully flips from $|\phi_{AB}^{-}\rangle$ to $|\phi_{AB}^{+}\rangle$ [Fig.~\ref{fig4} (c)]. Furthermore, as a reflection of $\Omega_{c}$, the braiding completeness can be manipulated through varying $E_{d}$ [see Fig.~\ref{fig4} (d)] \footnote{Notice that the braiding completeness in our results is not strictly follow the relation $|{\langle}\phi_{CD}^{+}(0)|\phi_{CD}^{-}(T_{s}){\rangle}|=t_{c}^{2}/(E_{d}^{2}+t_{c}^{2})$. Since the fermionic mode does not perfectly couple to the vMBSs, the effective $t_{c,{\rm eff}}$ is smaller than the $t_{c}$ set in the numerical simulation.}. Since only a single fermionic mode is required, our protocol brings experimental convenience in FeSCs platforms by driving a single quantum-dot structure with an AFM/STM tip to approach the two vortices alternately. Besides, since the braiding process induces a local charge transfer between vortices which is closely related to the braiding completeness \cite{liu_minimal_2021}, the braiding outcome may be readout by performing local charge sensing measurements through the AFM/STM tip \cite{berkovits_theory_2005,leoni_controlling_2011,gross_measuring_2009,miyahara_quantum_2017}.

	\textcolor[rgb]{0.00,0.00,1.00}{\emph{Conclusion.}}--We established the fermionic Y-junction to reflect the non-Abelian statistics of vMBSs onto the CFS of fermionic modes. We numerically demonstrated the effectiveness of the fermionic Y-junction in identifying vMBSs and their non-Abelian statistical properties. The dynamical error induced by the evolution of low-energy states is corrected through a Majorana version of the quantum Zeno effect. Moreover, we proposed a portable protocol to perform braiding operations over vMBSs using only a single fermionic mode. Our proposals will significantly simplify the experimental setup required for scalable TQC based on the FeSCs platforms.

%

	\begin{acknowledgments}
		We thank Hai-Wen Liu, Ying Jiang, and Qing-Feng Sun for fruitful discussion. This work is financially supported by the National Basic Research Program of China (Grants No. 2019YFA0308403), the National Natural Science Foundation of China (Grants No. 11974271 and No. 11822407), and the Strategic Priority Research Program of Chinese
		Academy of Sciences (Grant No. XDB28000000).
	\end{acknowledgments}
	
	\bibliography{draft_ming}

\end{document}


\title{Supplementary Materials for ``Unveiling non-Abelian statistics of vortex Majorana bound states in iron-based superconductors using fermionic modes''}

	\author{Ming Gong}
\affiliation{International Center for Quantum Materials, School of Physics, Peking University, Beijing 100871, China}

\author{Yijia Wu}
\affiliation{International Center for Quantum Materials, School of Physics, Peking University, Beijing 100871, China}

\author{Hua Jiang}
\affiliation{School of Physical Science and Technology and Institute for Advanced Study, Soochow University, Suzhou 215006, China.}

\author{Jie Liu}
\affiliation{Department of Applied Physics, School of Science, Xian Jiaotong University, Xian 710049, China}

\author{X. C. Xie}
\email{xcxie@pku.edu.cn}
\affiliation{International Center for Quantum Materials, School of Physics, Peking University, Beijing 100871, China}
\affiliation{Beijing Academy of Quantum Information Sciences, Beijing 100193, China}
\affiliation{CAS Center for Excellence in Topological Quantum Computation,
	University of Chinese Academy of Sciences, Beijing 100190, China}

\maketitle



\section{Hamiltonian for the surface topological superconducting system with a pair of vortices}
The Hamiltonian that holds a pair of vortices reads
\begin{eqnarray}
\scriptsize
\nonumber
H_{TSC}=&&\sum_{\mathbf{i}}\Big[\frac{i \hbar v_{F}}{2 a}(c_{\mathbf{i}}^{\dagger} \sigma_{y} c_{\mathbf{i}+\delta \hat{\mathbf{x}}}-c_{\mathbf{i}}^{\dagger} \sigma_{x} c_{\mathbf{i}+\delta \hat{\mathbf{y}}})-\frac{\mu}{2}c_{\mathbf{i}}^{\dagger}\sigma_{0} c_{\mathbf{i}}\nonumber\\
&&-\frac{W}{2a}(c_{\mathbf{i}}^{\dagger}\sigma_{z} c_{\mathbf{i}+\delta\hat{\mathbf{x}}}+c_{\mathbf{i}}^{\dagger} \sigma_{z}c_{\mathbf{i}+\delta \hat{\mathbf{y}}})+\frac{W}{a}c_{\mathbf{i}}^{\dagger}\sigma_{z} c_{\mathbf{i}}\nonumber\\
&&+\Delta tanh\frac{|\mathbf{i}-\mathbf{j_{1}}|}{\xi}    tanh\frac{|\mathbf{i}-\mathbf{j_{2}}|}{\xi}e^{i\theta(\mathbf{i}-\mathbf{j_{1}})-i\theta(\mathbf{i}-\mathbf{j_{2}})}c_{\mathbf{i},\uparrow}^{\dagger}c_{\mathbf{i},\downarrow}^{\dagger}\Big]+\mathrm{H.c.}
\label{eq:1}  
\end{eqnarray}
with $\mathbf{j_{1}}$ and $\mathbf{j_{2}}$ labeling the positions of the vortex and the anti-vortex \cite{fu_superconducting_2008,marchand_lattice_2012,gong_transport_2020}.

\begin{figure}[htbp]
	\includegraphics[width=0.8\textwidth]{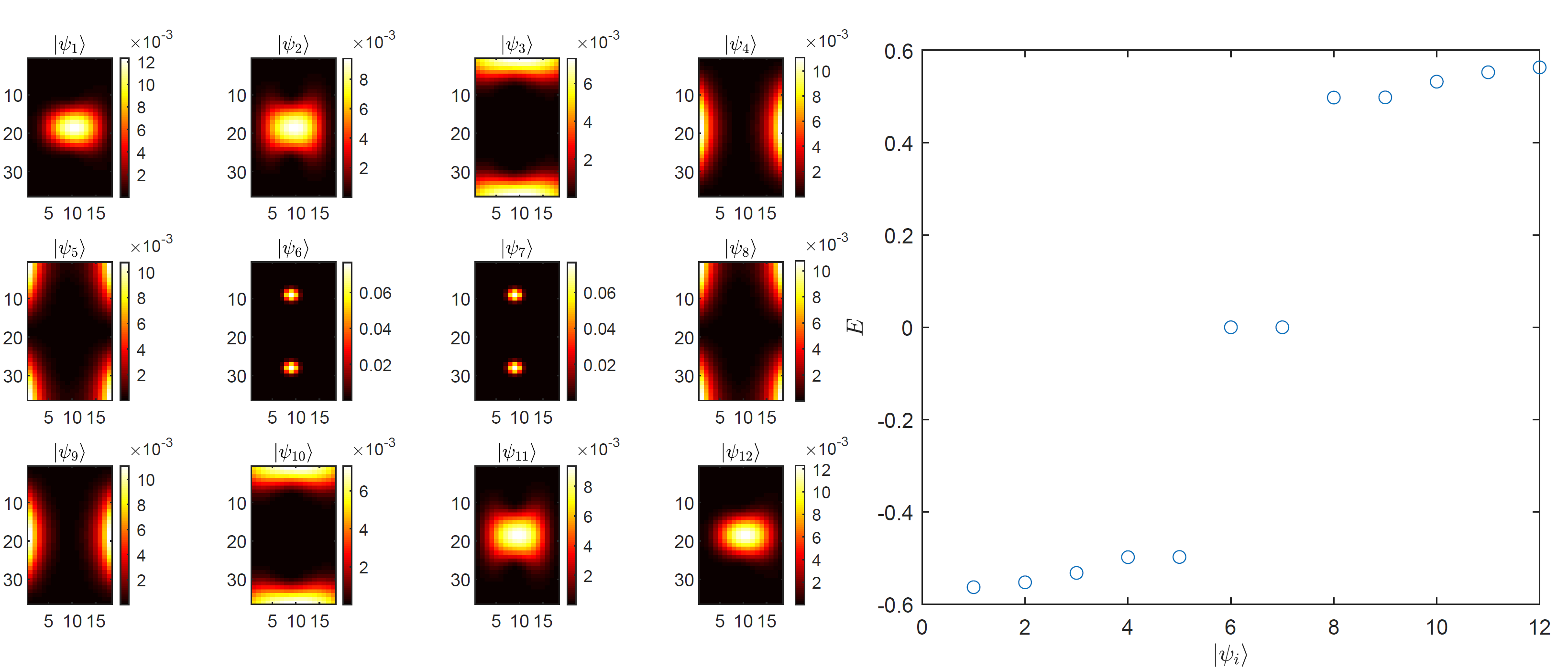}
	\caption{Spectrum and wave function distributions for the lowest 12 eigenstates. Here, $|\psi_{6}\rangle$ and $|\psi_{7}\rangle$ denote the unoccupation and occupation states of the fermionic mode encoded by vMBSs.}  
	\label{figs1}
\end{figure}

As mentioned in the main text, model parameters are taken as  $a=1$, $\hbar v_{F}=1$, $W=1$, $\mu=0$, $\Delta=1.5$ and $\xi=2$. In numerical simulations, the size of the system is taken as 36$\times$18 with the distance $L$=18 between vortices. To eliminate unnecessary edge states, we adopt the periodic boundary condition. We ignore the effect of the magnetic field by neglecting the vector potential $\mathbf{A}(\mathbf{r})$ in the Hamiltonian, since $\mathbf{B}$ is extremely weak for an extreme type II superconductor \cite{xu_topological_2016}. Fig. \ref{figs1} shows the spectrum and wave function distributions for the lowest 12 eigenstates for the Hamiltonian in Eq. \ref{eq:1}.  
\section{Non-Abelian braiding protocol for vortex Majorana bound states using single fermionic mode}
\begin{figure}[htbp]
	\includegraphics[width=0.9\textwidth]{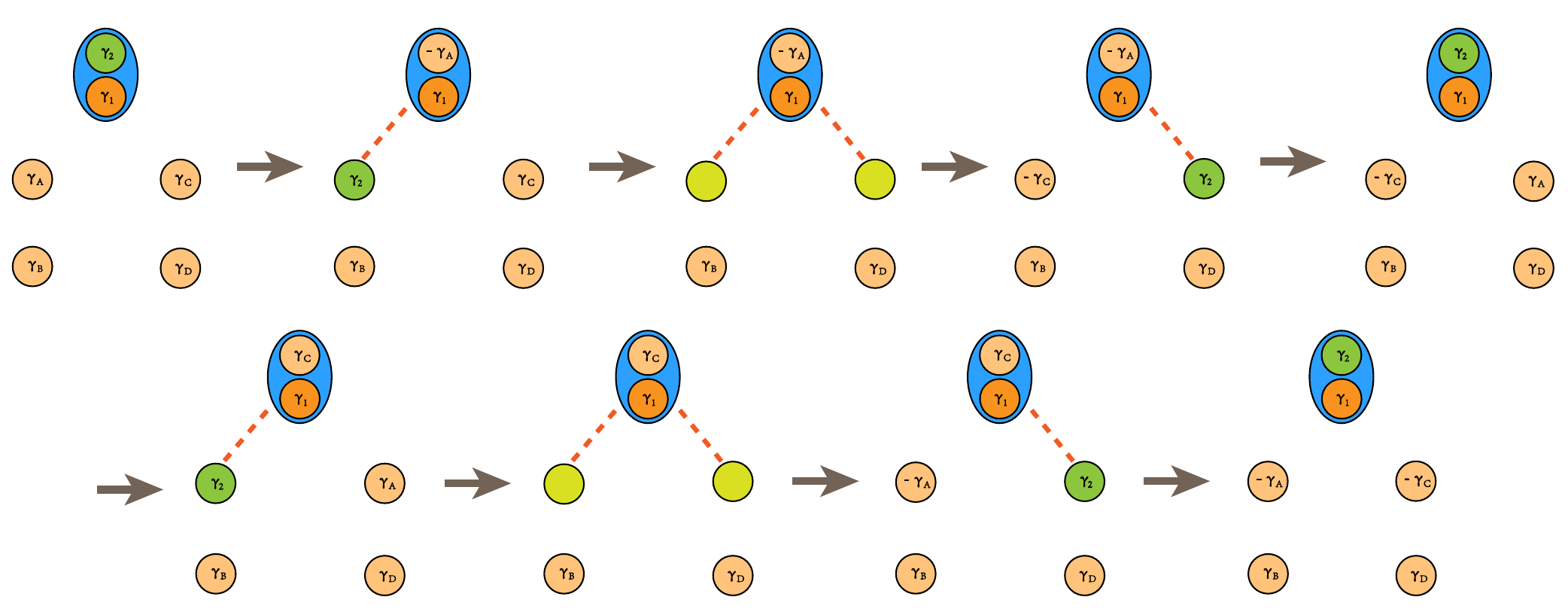}
	\caption{Illustration of the non-Abelian braiding protocol using only a single fermionic mode.}  
	\label{figs3}
\end{figure}

	The model Hamiltonian to realize our protocol is
\begin{eqnarray}
\nonumber
H_{M}=&&iE_{d}(\psi_{12}^\dagger\psi_{12}-\frac 12)+it_{1,A}\gamma_{1}\gamma_{A}/2+it_{1,C}\gamma_{1}\gamma_{C}/2\\
&&+i\epsilon_{AB}\gamma_{A}\gamma_{B}/2+i\epsilon_{CD}\gamma_{C}\gamma_{D}/2,
\end{eqnarray}
where $\psi_{12}=(\gamma_{1}+i\gamma_{2})/2$ is the fermionic mode encoded by Majorana modes $\gamma_{1}$ and $\gamma_{2}$ with energy $E_{d}$. $\gamma_{A}$, $\gamma_{B}$, $\gamma_{C}$ and $\gamma_{D}$ are vMBSs. $\epsilon_{AB}$ ($\epsilon_{CD}$) denotes the hybridization strength between $\gamma_{A}$ and $\gamma_{B}$ ($\gamma_{C}$ and $\gamma_{D}$), which can be neglected for weak hybridized vMBSs. The non-Abelian braiding operation is depicted in Fig. \ref{figs3} which is implemented through alternately turning on and off $t_{1,A}$ and $t_{1,C}$ \cite{liu_minimal_2021}. To successfully complete the braiding process, $t_{c}/E_{d}\gg1$ should be satisfied with $t_{c}$ the maximal value of $t_{1,A}(t_{1,C})$ during the operation.

In numerical simulations, the Hamiltonian that holds two pairs of vortices reads
\begin{eqnarray}
\scriptsize
\nonumber
H_{TSC}=&&\sum_{\mathbf{i}}\Big[\frac{i \hbar v_{F}}{2 a}(c_{\mathbf{i}}^{\dagger} \sigma_{y} c_{\mathbf{i}+\delta \hat{\mathbf{x}}}-c_{\mathbf{i}}^{\dagger} \sigma_{x} c_{\mathbf{i}+\delta \hat{\mathbf{y}}})-\frac{\mu}{2}c_{\mathbf{i}}^{\dagger}\sigma_{0} c_{\mathbf{i}}-\frac{W}{2a}(c_{\mathbf{i}}^{\dagger}\sigma_{z} c_{\mathbf{i}+\delta\hat{\mathbf{x}}}+c_{\mathbf{i}}^{\dagger} \sigma_{z}c_{\mathbf{i}+\delta \hat{\mathbf{y}}})+\frac{W}{a}c_{\mathbf{i}}^{\dagger}\sigma_{z} c_{\mathbf{i}}\nonumber\\
&&+\Delta tanh\frac{|\mathbf{i}-\mathbf{j_{1}}|}{\xi}    tanh\frac{|\mathbf{i}-\mathbf{j_{2}}|}{\xi}tanh\frac{|\mathbf{i}-\mathbf{k_{1}}|}{\xi}    tanh\frac{|\mathbf{i}-\mathbf{k_{2}}|}{\xi}e^{i\theta(\mathbf{i}-\mathbf{j_{1}})+i\theta(\mathbf{i}-\mathbf{k_{1}})-i\theta(\mathbf{i}-\mathbf{j_{2}})-i\theta(\mathbf{i}-\mathbf{k_{2}})}c_{\mathbf{i},\uparrow}^{\dagger}c_{\mathbf{i},\downarrow}^{\dagger}\Big]+\mathrm{H.c.}
\label{eq:2}  
\end{eqnarray}
with $\mathbf{j_{1}}$ ($\mathbf{k_{1}}$) and $\mathbf{j_{2}}$ ($\mathbf{k_{2}}$) labeling the positions of the vortex and the anti-vortex. Parameters are same to Eq. \ref{eq:1}. The system size is taken as 52$\times$52 with $L$=26. 
	\begin{figure}[htbp]
	\includegraphics[width=0.7\textwidth]{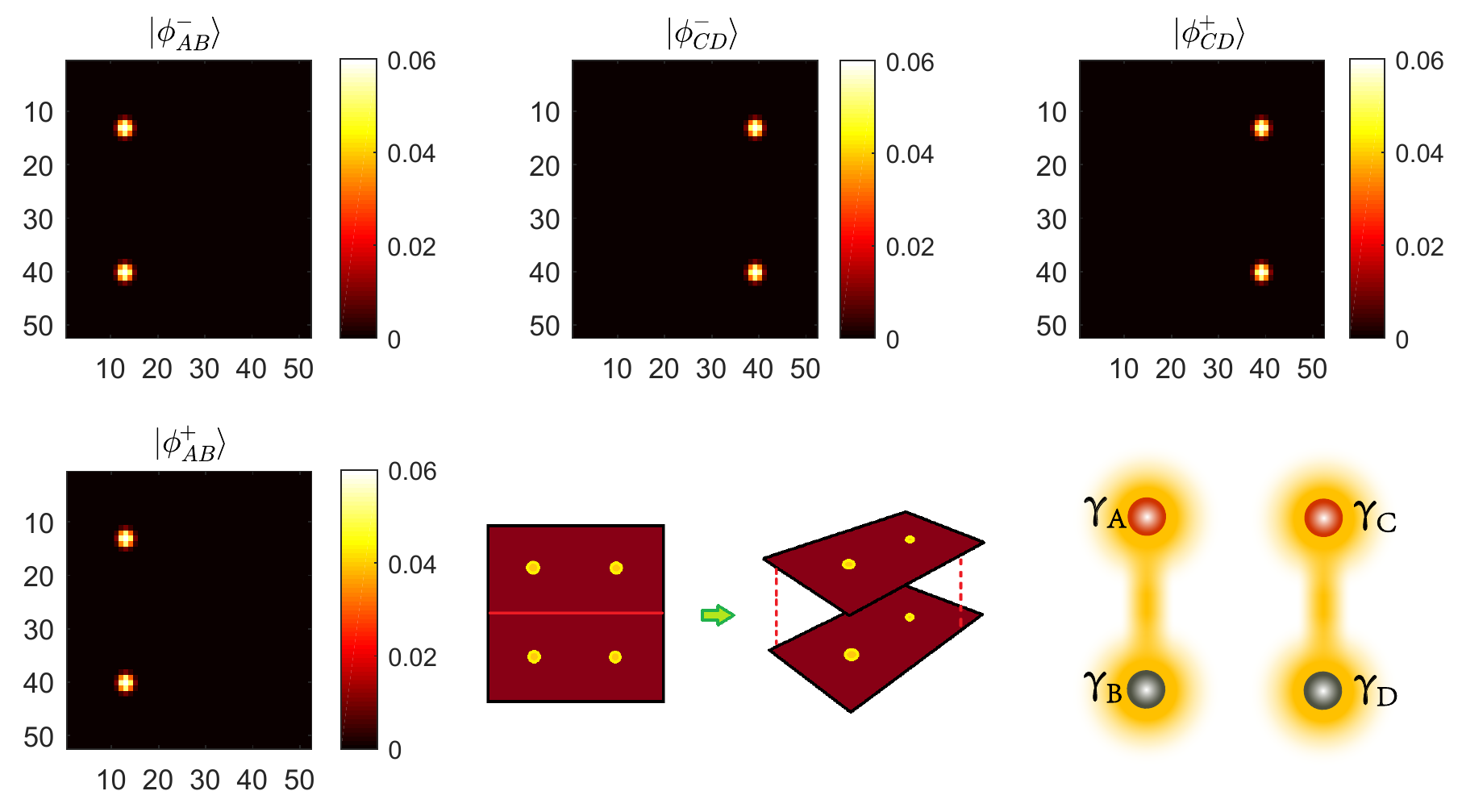}
	\caption{Wave function distributions for the occupation and unoccupation states of fermionic mode $\phi_{AB}$ ($\phi_{CD}$) encoded by $\gamma_{A}$ and $\gamma_{B}$ ($\gamma_{C}$ and $\gamma_{D}$).}  
	\label{figs2}
\end{figure}

The Majorana bound state on the vortex core can be decomposed into the superposition of fermion modes:
\begin{equation}
\gamma_{i}=c+c^{\dagger}
\end{equation}
where $c^{\dagger}(c)$ contains only particle$($hole$)$ degree of freedom, which is localized near the vortex core. When a spin-polarized (say, spin up state labeled by $\uparrow$) fermionic mode is coupled to the vortex core modes, the coupling Hamiltonian reads
	\begin{equation}
	H_{d}=E_{d}d^{\dagger}_{\uparrow}d_{\uparrow}+td^{\dagger}_{\uparrow}c_{\uparrow}+t^{*}c_{\uparrow}^{\dagger}d_{\uparrow}
	\end{equation}
where $E_{d}$ is the single-particle energy of the fermionic mode and $t$ is the effective coupling strength between the fermionic mode and the vortex core. The total Hamiltonian $H$ of the system is
	\begin{equation}
	H_{total}=H_{TSC}+E_{d}d^{\dagger}_{\uparrow}d_{\uparrow}+H_{couple}
	\end{equation}
where
	\begin{equation}
	H_{couple}=t_{1}d^{\dagger}_{\uparrow}c_{1\uparrow}+t_{2}d^{\dagger}_{\uparrow}c_{2\uparrow}+h.c.
	\end{equation}	
In our model, $t_{1}$ and $t_{2}$ satisfy $t_{1}/|t_{1}|=\pm it_{2}/|t_{2}|$. It is because the naive phase choice of $\Delta(\mathbf{i})$ breaks the translational symmetry of the vortex lattice, leading to a $\pi$ phase difference between the nearest vortices \cite{pathak_majorana_2021-1}. Such a consequence originates from the naive omitting of the vector potential $\mathbf{A}(\mathbf{r})$ in $H_{TSC}$. In real systems, such a phase difference does not exist because that the introduction of $\mathbf{A}(\mathbf{r})$ will exactly cancel such a phase (each vortex core in a type-II superconductor is surrounded by a half-quantized magnetic flux with radius $\lambda$ the penetration depth), making the system translational invariant. Therefore, we introduce an additional $\pm \pi/2$ phase to the electron degree of freedom in our model, which manifests a phase difference $t_{1}/|t_{1}|=\pm it_{2}/|t_{2}|$ between the coupling strengths.

	\bibliography{draft_ming}